\documentclass{article}
\usepackage{spconf,amsmath,amssymb,graphicx}
\usepackage{booktabs}

\newcommand{\knn}{$k$NN}
\newcommand{\zmin}{\textsc{z-min}}

\title{Scoring Backends Matter More Than Pooling: A Systematic Study of Training-Free Anomalous Sound Detection under Domain Shift}

\name{Jingwen Zhou \qquad Mingzhe Wang}
\address{Xidian University, Xi'an, China}

\begin{document}
\ninept
\maketitle

\begin{abstract}
Training-free anomalous sound detection (ASD) scores a test clip against a memory bank of normal embeddings from a frozen pretrained audio encoder. Recent work attributes domain-shift robustness mainly to how frame-level features are \emph{pooled} over time; the \emph{scoring backend} applied on top of the pooled embedding has received far less systematic attention. Using a single frozen BEATs encoder on the DCASE~2023 Task~2 development set (all seven machine types), we cross four classical backends---nearest-neighbor cosine distance, Mahalanobis distance, locally density-normalized $k$NN, and PCA-subspace reconstruction residual---with three temporal poolings (mean, GeM, max). Switching the backend moves target-domain AUC by 13.8 points on average (up to 53.8), whereas switching the pooling moves it by only 3.2 points: in this training-free regime, the backend, not the pooling, dominates domain-shift robustness. No backend wins everywhere, but the machine-dependent pattern reproduces on the DCASE~2025 development data (fan, bearing). Exploiting this, we propose a label-free score fusion that z-normalizes each backend with its training-bank self-scores and takes the minimum; it reaches a harmonic-mean target AUC of 63.3\% versus 64.4\% for the per-machine oracle, surpassing every fixed single backend while preserving source-domain accuracy. We also report a negative result: selecting a backend by source-domain pseudo-validation with proxy outliers fails, because all backends saturate on the proxy task.
\end{abstract}


\section{Introduction}
\label{sec:intro}

Unsupervised ASD for machine condition monitoring, as standardized by the DCASE Task~2 series \cite{koizumi2020dcase,dohi2022dcase,dohi2023dcase,nishida2025dcase}, asks a system to detect anomalous sounds having heard only \emph{normal} clips of a machine. Since 2022 the task has emphasized \emph{domain shift}: the bank of normal training clips is dominated by a source domain (990 clips), while only 10 clips cover a target domain whose operating or recording conditions changed; systems are scored separately on source- and target-domain test clips.

A strong and increasingly popular recipe is entirely \emph{training-free} \cite{wilkinghoff2023pretrained,pooling2026}: embed all normal clips with a frozen pretrained audio encoder---BEATs \cite{chen2023beats} or alternatives such as wav2vec~2.0, HuBERT, AST, and PANNs \cite{baevski2020wav2vec2,hsu2021hubert,gong2021ast,kong2020panns}---embed the test clip, and convert distances to the normal bank into an anomaly score. This mirrors memory-bank methods in industrial visual inspection: $k$NN on frozen deep features \cite{bergman2020dn2}, Gaussian modeling of pretrained features \cite{rippel2021gaussian,defard2021padim}, and their modern descendants PatchCore \cite{roth2022patchcore}, AnomalyDINO \cite{damm2024anomalydino}, and MuSc \cite{li2024musc}. It also contrasts with trained ASD systems that fine-tune pretrained models \cite{jiang2024anopatch} or learn discriminative embeddings through angular-margin auxiliary tasks \cite{wilkinghoff2023design}. In the audio domain, recent work argues that adaptive temporal pooling can be a critical design choice in the training-free pipeline \cite{pooling2026}. Our study is deliberately orthogonal: it holds the pooling family to standard static choices (mean, GeM, max) and asks how much robustness is determined by the scoring rule placed on top.

This paper examines the complementary axis that the pooling-centric view leaves fixed: the \emph{scoring backend}. Classical anomaly scores differ in what they assume about the normal manifold---local neighborhoods (\knn{} \cite{ramaswamy2000}), a global Gaussian (Mahalanobis \cite{mahalanobis1936}), local density ratios (LOF-style normalization \cite{breunig2000lof}), or a linear subspace (PCA residual \cite{hoffmann2007kpca}). Under domain shift these assumptions are stressed in different ways, so backends should not be interchangeable. We ask: \emph{with the backbone, pooling, and bank held fixed, how much of domain-shift robustness is decided by the scoring backend alone?}

Our contributions are:
\begin{itemize}
\item \textbf{A controlled backend\,$\times$\,pooling study.} On all seven DCASE~2023 Task~2 development machines, with one frozen BEATs encoder, we cross four backends with three poolings (84 configurations). Backend choice moves target-domain AUC by $13.8$ points on average (median $8.4$, max $53.8$); pooling choice moves it by $3.2$ points (Sec.~\ref{ssec:swing}). In this frozen-backbone, clip-level regime the backend is the dominant lever, $\sim$4--6$\times$ over pooling, complementing the pooling-centric analysis of \cite{pooling2026}.
\item \textbf{No free lunch, but a stable pattern.} No backend wins on all machines; density-normalized \knn{} is best on the target domain for 4/7 machines and PCA residual beats plain \knn{} on 3/7. The qualitative pattern---including a striking source/target inversion of density-normalized \knn{} on \emph{fan}---reproduces on the independent DCASE~2025 development data (Sec.~\ref{ssec:crossyear}).
\item \textbf{A label-free fusion that nearly closes the oracle gap.} Z-normalizing each backend by its \emph{training-bank self-scores} and taking the \emph{minimum} achieves 63.3\% harmonic-mean target AUC versus 64.4\% for the per-machine oracle and 55.5\% for the expected blind pick, without sacrificing the source domain (Sec.~\ref{ssec:fusion}).
\item \textbf{An honest negative result.} Backend selection by source-domain pseudo-validation with proxy outliers (other machines' embeddings) fails: all backends saturate near pseudo-AUC $1.0$, leaving no selection signal (Sec.~\ref{ssec:fusion}).
\end{itemize}

\section{Method}
\label{sec:method}

\subsection{Training-free ASD pipeline}
\label{ssec:pipeline}
For each machine type we are given $N{=}1{,}000$ normal training clips ($990$ source, $10$ target). A frozen encoder maps a clip $x$ to frame features $F(x)\in\mathbb{R}^{T\times d}$; a temporal pooling $\phi$ produces the clip embedding $\mathbf{e}=\phi(F(x))\in\mathbb{R}^{d}$. The bank is $\mathcal{B}=\{\mathbf{e}_1,\dots,\mathbf{e}_N\}$. At test time, a backend $s(\cdot)$ maps a test embedding to a scalar anomaly score. We use BEATs\_iter3+\,(AS2M, self-supervised) \cite{chen2023beats}, $d{=}768$, and study three poolings: \emph{mean}, \emph{GeM} with $p{=}3$ \cite{radenovic2019gem} applied to activations clamped at $10^{-6}$, and \emph{max}.

\subsection{Four scoring backends}
\label{ssec:backends}
Let $d_c(\mathbf{a},\mathbf{b})=1-\frac{\mathbf{a}^{\top}\mathbf{b}}{\lVert\mathbf{a}\rVert\,\lVert\mathbf{b}\rVert}$ denote cosine distance.

\textbf{(a) Nearest-neighbor cosine (\knn{}).}
$s_{\mathrm{knn}}(\mathbf{e})=\min_{i} d_c(\mathbf{e},\mathbf{e}_i)$, the classical distance-based outlier score \cite{ramaswamy2000} and standard memory-bank rule on deep features \cite{bergman2020dn2,roth2022patchcore}.

\textbf{(b) Mahalanobis.} With bank mean $\boldsymbol{\mu}$ and a Ledoit--Wolf shrinkage covariance $\hat{\Sigma}$ \cite{ledoit2004} (necessary since $d{=}768\approx N$),
$s_{\mathrm{mah}}(\mathbf{e})=\sqrt{(\mathbf{e}-\boldsymbol{\mu})^{\top}\hat{\Sigma}^{-1}(\mathbf{e}-\boldsymbol{\mu})}$ \cite{mahalanobis1936}, i.e., Gaussian modeling of pretrained features as in \cite{rippel2021gaussian,defard2021padim}.

\textbf{(c) Locally density-normalized \knn{}.} Inspired by LOF \cite{breunig2000lof}: with $i^{*}$ the nearest bank index and $\rho_i$ the mean cosine distance of bank item $i$ to its $k{=}5$ nearest bank neighbors (self excluded),
$s_{\mathrm{lnorm}}(\mathbf{e})= s_{\mathrm{knn}}(\mathbf{e})/\rho_{i^{*}}$.
The normalization discounts distances in sparse regions of the bank---e.g., the 10 target-domain exemplars. Closely related local-density score rescalings were recently shown to be state-of-the-art for domain-generalized ASD \cite{wilkinghoff2025localdensity}.

\textbf{(d) PCA-subspace residual.} Fit PCA on the bank retaining 90\% variance ($P\in\mathbb{R}^{d\times r}$, mean $\boldsymbol{\mu}$); score by the reconstruction residual
$s_{\mathrm{pca}}(\mathbf{e})=\lVert(\mathbf{e}-\boldsymbol{\mu})-PP^{\top}(\mathbf{e}-\boldsymbol{\mu})\rVert_2$,
i.e., the energy outside the normal subspace \cite{hoffmann2007kpca}. With mean pooling, $r$ ranges from 11 (fan) to 59 (ToyTrain).

All four backends are deterministic, training-free, and run in seconds per machine on CPU.

\subsection{Label-free combination}
\label{ssec:combine}
\textbf{Self-score z-normalization and fusion.} Backend scores live on incommensurable scales, so fusion first requires score normalization \cite{jain2005score}. We calibrate each backend with its own \emph{bank self-scores}: every bank item is scored against the bank (leave-one-out for the \knn{}-type backends, i.e., with the diagonal excluded from the nearest-neighbor search; direct scoring for Mahalanobis and PCA), giving a mean $\mu_b$ and standard deviation $\sigma_b$ per backend $b$; a test score is standardized as $z_b=(s_b-\mu_b)/\sigma_b$. This uses no test-set statistics and no labels. We fuse by the classical combination rules \cite{kittler1998combining}: $\operatorname{mean}_b z_b$, $\max_b z_b$, or $\min_b z_b$. Whereas outlier ensembles typically average or maximize \cite{aggarwal2015ensembles}, the \emph{minimum} is deliberately conservative: a clip is anomalous only if \emph{every} view of the normal manifold rejects it. Under domain shift, individual backends produce false alarms on target-domain normals; \zmin{} suppresses exactly these (Sec.~\ref{ssec:fusion}).

\textbf{Pseudo-validation backend selection (negative control).} We also test whether the best backend can be \emph{selected} without labels: hold out source normals (5 folds), score them against the remaining bank, use 500 randomly drawn bank embeddings of the \emph{other} machines as proxy outliers, and pick the backend with the highest held-out-vs-proxy AUC (averaged over folds; fixed seed). This mimics proxy-outlier validation and is fully label-free.

\section{Experiments}
\label{sec:exp}

\textbf{Setup.} DCASE~2023 Task~2 development set \cite{dohi2023dcase} (ToyADMOS2 \cite{harada2021toyadmos2} and MIMII~DG \cite{dohi2022mimii} recordings), all seven machines; per machine, $1{,}000$ normal training clips and $200$ test clips ($50$ normal${+}50$ anomalous per domain). Audio is 16\,kHz; clips are 10\,s (ToyCar/ToyTrain: first 10 of 12\,s). Following the official protocol \cite{dohi2023dcase}, AUC is computed per domain using that domain's normal clips against \emph{all} anomalous clips of the section, and pAUC ($p{=}0.1$) over the full section. For cross-year validation we use the DCASE~2025 development data \cite{nishida2025dcase}, restricted to two machines (fan, bearing); we verified the 2025 audio differs from 2023 (near-zero waveform correlation at matched indices). All results are from single deterministic runs (no trainable parameters; selection seed fixed).

\begin{figure}[t]
\centering
\includegraphics[width=\columnwidth]{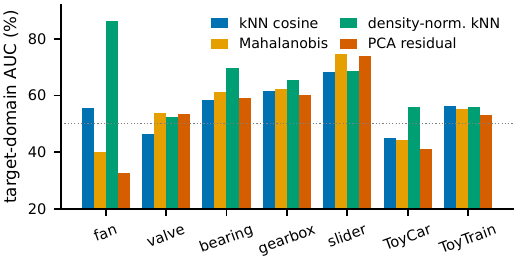}
\caption{Target-domain AUC of the four scoring backends on all seven DCASE2023 development machines (frozen BEATs, mean pooling; dotted line = chance). The backend changes the outcome drastically and no backend wins everywhere.}
\label{fig:target}
\end{figure}

\subsection{Main table: machines $\times$ backends}
\label{ssec:main}
Table~\ref{tab:main} reports the full matrix with mean pooling (visualized in Fig.~\ref{fig:target}), alongside the official autoencoder baseline (AE-MSE \cite{harada2023firstshot}; averages over five runs from \cite{dohi2023dcase}). Three observations. (i) \emph{Source-domain} AUC of the plain \knn{}, Mahalanobis, and PCA backends is competitive---never below the AE baseline on any machine---confirming that a frozen BEATs bank is a strong substrate. (ii) \emph{Target-domain} AUC varies wildly with the backend: on fan it spans $32.6$ (PCA) to $86.3$ (density-normalized \knn{}). (iii) Density-normalized \knn{} is the best target-domain backend on 4/7 machines, at the price of a severe \emph{source} collapse on fan ($32.4$): normalizing by local bank density over-compensates exactly when the source cluster is tight---a phenomenon we observe again in 2025 (Sec.~\ref{ssec:crossyear}).

\textbf{Why fan swings from 32.6 to 86.3.} The fan bank is a geometric outlier consistent with this extreme behavior: 90\% of its variance is captured by only $r{=}11$ principal components, versus $r{\geq}40$ for every other machine---its source cluster is unusually compact and low-dimensional. The two extreme backends respond to this same geometry with opposite signs. PCA residual fits the normal subspace to the source-dominated bank, so shifted target normals fall outside it and are flagged: the machine's highest source AUC ($89.9$) coexists with a near-pathological target AUC ($32.6$). Density-normalized \knn{} divides by local bank density: inside the tight source cluster $\rho_{i^{*}}$ is small, so even mild deviations of source-domain normals are amplified into false alarms (source $32.4$), whereas around the 10 sparse target exemplars $\rho_{i^{*}}$ is large and target normals are forgiven (target $86.3$). The $53.8$-point swing is therefore not noise but two coherent, mutually inverse responses to one compact source manifold; the remaining backends interpolate between the extremes (\knn{} $55.7$, Mahalanobis $40.1$ on the target domain).

\begin{table*}[t]
\centering
\caption{DCASE 2023 Task 2 development set, all seven machines: AUC$_\mathrm{source}$ / AUC$_\mathrm{target}$ / pAUC (\%), frozen BEATs, mean pooling. Official AE-MSE baseline \cite{harada2023firstshot} as reported in \cite{dohi2023dcase}. \textbf{Bold} = best target-domain AUC per machine.}
\label{tab:main}
\begin{tabular}{lccccc}
\toprule
Machine & \knn{} cosine & Mahalanobis & density-norm.\ \knn{} & PCA residual & AE-MSE baseline \cite{dohi2023dcase} \\
\midrule
fan      & 86.20 / 55.70 / 59.00 & 86.02 / 40.12 / 54.16 & 32.42 / \textbf{86.32} / 53.53 & 89.88 / 32.56 / 51.26 & 80.19 / 36.18 / 59.04 \\
valve    & 62.50 / 46.50 / 52.53 & 63.02 / \textbf{53.92} / 53.58 & 58.82 / 52.36 / 49.42 & 63.00 / 53.48 / 53.42 & 55.35 / 50.69 / 51.18 \\
bearing  & 71.44 / 58.46 / 56.32 & 72.56 / 61.16 / 54.63 & 68.64 / \textbf{69.80} / 60.84 & 71.78 / 58.98 / 53.89 & 65.92 / 55.75 / 50.42 \\
gearbox  & 73.14 / 61.60 / 56.84 & 74.18 / 62.24 / 58.89 & 67.22 / \textbf{65.54} / 56.00 & 68.60 / 60.12 / 53.74 & 60.31 / 60.69 / 53.22 \\
slider   & 82.94 / 68.38 / 55.26 & 89.44 / \textbf{74.60} / 56.00 & 78.70 / 68.60 / 62.21 & 88.78 / 73.76 / 54.89 & 70.31 / 48.77 / 56.37 \\
ToyCar   & 75.36 / 44.98 / 51.32 & 70.88 / 44.36 / 52.11 & 64.20 / \textbf{55.92} / 50.89 & 72.90 / 41.02 / 51.26 & 70.10 / 46.89 / 52.47 \\
ToyTrain & 73.10 / \textbf{56.36} / 49.16 & 80.28 / 55.10 / 49.47 & 73.64 / 55.92 / 51.05 & 79.98 / 53.24 / 49.47 & 57.93 / 57.02 / 48.57 \\
\bottomrule
\end{tabular}
\end{table*}

\begin{figure}[t]
\centering
\includegraphics[width=\columnwidth]{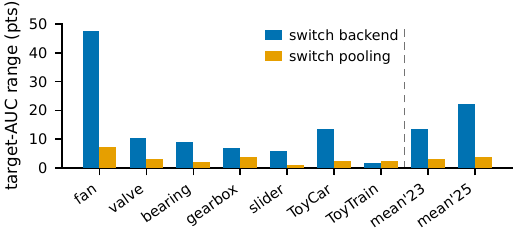}
\caption{Per-machine range of target-domain AUC (DCASE2023, frozen BEATs) when switching the scoring backend (pooling fixed) vs.\ switching the temporal pooling (backend fixed); rightmost bars: averages for DCASE2023 (7 machines) and DCASE2025 (fan, bearing).}
\label{fig:swing}
\end{figure}

\subsection{Backend vs.\ pooling: which lever moves domain robustness?}
\label{ssec:swing}
For every (machine, pooling) cell we compute the \emph{range} of target-domain AUC across the four backends, and symmetrically, for every (machine, backend) cell the range across the three poolings. Table~\ref{tab:swing} summarizes. Switching the backend moves target AUC by $13.75$ points on average---over four times the $3.23$ points moved by switching the pooling; the same holds on the source domain ($13.58$ vs.\ $2.97$) and on the 2025 data ($22.62$ vs.\ $4.01$, $5.6\times$). Per machine, the backend range exceeds the pooling range on 6/7 machines (the exception, ToyTrain, has both small). The largest pooling-induced range for a fixed backend is $15.9$ points (PCA residual on fan); yet the per-machine best backend is unchanged by pooling on 4/7 machines, and where it changes (valve, slider, ToyTrain) the decision is mostly a near-tie: the winner leads the runner-up by less than one point in eight of the nine affected (machine, pooling) cells, the single exception being valve with max pooling, where density-normalized \knn{} leads by 8.5 points. The best target value of the entire $84$-configuration grid ($86.3$, fan) is set by the backend choice (density-normalized \knn{}), not by the pooling. \emph{In this clip-level training-free regime, the scoring backend---not the temporal pooling---is the dominant design choice for domain robustness.} Our grid covers static poolings; the adaptive deviation-based pooling of \cite{pooling2026} is a complementary axis, and crossing it with the backends studied here is a natural next step. Fig.~\ref{fig:swing} visualizes the asymmetry.

Per machine, the asymmetry is sharpest exactly where domain shift hurts most. The mean backend-induced range is $48.2$ points on fan and $13.6$ on ToyCar, against pooling-induced ranges of $7.3$ and $2.6$; at the other end, ToyTrain compresses both ranges ($1.9$ vs.\ $2.4$)---all twelve of its configurations land within five points of target AUC ($53.2$--$58.3$), so neither lever matters and the frozen embedding itself appears to be the bottleneck there. The two largest pooling-induced ranges are also instructive: both arise where max pooling partially rescues an otherwise failing backend (PCA residual on fan, $32.6{\rightarrow}48.4$; density-normalized \knn{} on valve, $52.4{\rightarrow}61.1$, the valve/max exception above). Pooling thus acts mainly through its \emph{interaction} with a particular backend rather than by uniformly lifting all backends---which is precisely why treating the backend as the free variable exposes the larger effect.

\begin{table}[t]
\centering
\caption{Mean (median / max) range of target-domain AUC (points) induced by switching one factor while holding the other fixed.}
\label{tab:swing}
\footnotesize\setlength{\tabcolsep}{3pt}
\begin{tabular}{lcc}
\toprule
Data & switch backend & switch pooling \\
\midrule
2023 (7 mach.) & \textbf{13.75} (8.36 / 53.76) & 3.23 (2.30 / 15.88) \\
2025 (2 mach.) & \textbf{22.62} & 4.01 \\
\bottomrule
\end{tabular}
\end{table}

\subsection{Cross-year validation (DCASE 2025)}
\label{ssec:crossyear}
Table~\ref{tab:y2025} repeats the study on the DCASE~2025 development data for the two machines fan and bearing (mean pooling); conclusions for the remaining machines are left to future work. All three qualitative findings reproduce on audio recorded for a different edition: (i) the backend swing dwarfs the pooling swing (Table~\ref{tab:swing}); (ii) the fan source/target inversion of density-normalized \knn{} recurs ($33.7$ source vs.\ $65.5$ target); (iii) PCA residual again beats plain \knn{} on bearing's target domain (${+}2.6$) and again fails on fan. The machine-dependent backend pattern is a stable property, not a 2023 artifact.

\begin{table}[t]
\centering
\caption{DCASE 2025 development data, two machines (fan, bearing): AUC$_\mathrm{source}$ / AUC$_\mathrm{target}$ (\%), frozen BEATs, mean pooling.}
\label{tab:y2025}
\footnotesize\setlength{\tabcolsep}{2pt}
\begin{tabular}{lcccc}
\toprule
Machine & \knn{} & Mahal. & d.-n.\ \knn{} & PCA res. \\
\midrule
fan     & 61.20 / 41.20 & 78.30 / 26.52 & 33.72 / \textbf{65.46} & 77.16 / 24.82 \\
bearing & 60.58 / 48.02 & 65.92 / 53.74 & 66.38 / \textbf{57.04} & 66.18 / 50.64 \\
\bottomrule
\end{tabular}
\end{table}

\begin{figure}[t]
\centering
\includegraphics[width=\columnwidth]{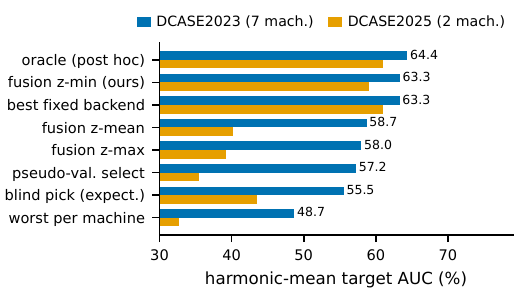}
\caption{Label-free combinations vs.\ reference points: harmonic-mean target-domain AUC across machines on DCASE2023 (7 machines) and DCASE2025 (fan, bearing). \zmin{} fusion nearly closes the gap to the per-machine oracle on both benchmark editions.}
\label{fig:fusion}
\end{figure}

\subsection{Label-free fusion and the oracle gap}
\label{ssec:fusion}
Table~\ref{tab:fusion} and Fig.~\ref{fig:fusion} evaluate the combinations of Sec.~\ref{ssec:combine} by the harmonic mean of target-domain AUC across machines (the DCASE-style aggregate). \zmin{} fusion reaches $63.32\%$ on 2023---within $1.05$ points of the per-machine oracle ($64.37\%$), $7.8$ points above the expected blind pick, and above the best \emph{fixed} backend (density-normalized \knn{}, $63.29\%$). Crucially, unlike that backend, \zmin{} does not sacrifice the source domain: source harmonic mean $69.61\%$ vs.\ $58.75\%$, and the official-style aggregate over AUC$_\mathrm{s}$, AUC$_\mathrm{t}$ and pAUC of all machines is $62.41\%$ vs.\ $58.61\%$. On 2025, \zmin{} is again within $2.0$ points of the oracle (fan $62.7$ vs.\ $65.5$, bearing $55.8$ vs.\ $57.0$) and ranks first on the official-style aggregate ($53.97\%$). On fan, \zmin{} attains the highest pAUC of the entire study ($70.21\%$).

Per machine, \zmin{} is the best of the three fusion rules on 6/7 of the 2023 machines and tracks the oracle closely; on gearbox ($66.7$ vs.\ $65.5$) and ToyTrain ($58.6$ vs.\ $56.4$) it even \emph{exceeds} the best single backend---something a per-machine selection, oracle included, cannot do. The one exception is slider ($71.2$ vs.\ $75.6$ for z-mean): there every backend already works (worst target AUC $68.4$), so there are few false alarms to veto and the conservative minimum only discards evidence. This delineates when \zmin{} pays off: precisely where backends disagree on target normals. Fan is the extreme case---z-mean ($65.8$) is dragged down by the two backends that misfire there (Mahalanobis $40.1$, PCA $32.6$), while \zmin{} recovers $85.4$, within one point of the oracle, and simultaneously lifts source AUC to $58.1$ from the $32.4$ of its best member. True anomalies, in contrast, survive the minimum: each backend is individually competent on the source domain (Table~\ref{tab:main}), so an anomalous clip is far from the bank under \emph{every} view and keeps a high fused score. This is the design intuition made quantitative: domain shift inflates \emph{some} backends' scores on target normals; requiring consensus to reject suppresses exactly these false alarms.

\textbf{Negative result: pseudo-validation selection.} The proxy-outlier selector is uninformative: held-out source normals versus other-machine proxies yields pseudo-AUC $\in[0.965,1.0]$ for \emph{all} backends on \emph{all} machines---other machines are simply too far from the bank---so the argmax degenerates to noise. The selected backends score $57.24\%$ (2023; barely above blind pick) and $35.51\%$ (2025; below blind pick). The failure mode is itself informative: across 2023 the selector never picks density-normalized \knn{}, the best fixed backend, but oscillates between plain \knn{} (3 machines) and Mahalanobis (4); on 2025 it picks Mahalanobis for both machines, including fan, where Mahalanobis has the \emph{worst} target AUC ($26.5$)---hence the below-blind-pick aggregate. Discriminating between backends would require \emph{near-manifold} proxies whose pseudo-AUCs do not saturate: plausible candidates are perturbed or interpolated same-machine embeddings, or, in the first-shot setting \cite{harada2023firstshot}, leave-one-out scoring of the 10 target exemplars themselves. We leave this to future work.

\begin{table}[t]
\centering
\caption{Harmonic mean of target-domain AUC (\%) across machines. Oracle picks the best backend per machine post hoc.}
\label{tab:fusion}
\footnotesize\setlength{\tabcolsep}{3pt}
\begin{tabular}{lcc}
\toprule
Method & 2023 (7 mach.) & 2025 (2 mach.) \\
\midrule
oracle (per machine, post hoc) & 64.37 & 60.96 \\
\textbf{fusion \zmin{} (ours)} & \textbf{63.32} & \textbf{59.01} \\
best fixed backend (d.-n.\ \knn{}) & 63.29 & 60.96 \\
fusion z-mean & 58.71 & 40.21 \\
fusion z-max & 57.98 & 39.27 \\
pseudo-validation selection & 57.24 & 35.51 \\
expected blind pick & 55.55 & 43.53 \\
worst per machine & 48.70 & 32.73 \\
\bottomrule
\end{tabular}
\end{table}

\section{Conclusion}
\label{sec:conclusion}
With backbone, bank, and pooling held fixed, the scoring backend is the dominant---and underexamined---design choice for domain-robust training-free ASD with a frozen BEATs backbone: it moves target-domain AUC 4--6$\times$ more than temporal pooling, its machine-dependent strengths are stable across benchmark editions, and a simple training-bank-calibrated \zmin{} fusion captures most of the per-machine oracle without labels, extra data, or training.

Limitations remain. The study is clip-level: frame- or patch-level memory banks in the style of PatchCore \cite{roth2022patchcore} may interact differently with the backends, and the adaptive pooling of \cite{pooling2026} is outside our grid. All conclusions are conditioned on a single backbone; encoders trained with different objectives may reshape the bank geometry that, as the fan analysis shows, drives backend behavior. Backend hyperparameters were fixed a priori ($k{=}5$ local neighbors, $90\%$ retained variance, GeM $p{=}3$)---consistent with the training-free setting, but leaving their sensitivity unmeasured. The cross-year check covers two machines of the 2025 development data. Finally, our label-free backend \emph{selection} attempt failed; we report it to delimit what source-domain pseudo-validation with far-from-manifold proxies can and cannot do.

\bibliographystyle{IEEEbib}
\bibliography{refs}

\end{document}